\documentclass[twocolumn]{aastex631}

\usepackage{xspace}
\usepackage{multirow}

\newcommand{\SNRUV}{3.6\xspace}
\newcommand{\SNRG}{4.5\xspace}
\newcommand{\SNRR}{5.8\xspace}

\begin{document}

\title{UV-Optical Emission of AB Aur b is Consistent with Scattered Stellar Light}
\shorttitle{A UV-Optical SED of AB Aur b}

\correspondingauthor{Yifan Zhou}
\email{yzhou@virginia.edu}
\author[0000-0003-2969-6040]{Yifan Zhou}
\affiliation{Department of Astronomy, The University of Texas at Austin,  2515 Speedway, Stop C1400 Austin, TX 78712, USA}
\affiliation{Department of Astronomy, The University of Virginia, 530 McCormick Rd, Charlottesville, VA 22904, USA}

\author[0000-0003-2649-2288]{Brendan P. Bowler}
\affiliation{Department of Astronomy, The University of Texas at Austin,  2515 Speedway, Stop C1400 Austin, TX 78712, USA}

\author[0000-0002-8537-6669]{Haifeng Yang}
\affiliation{Kavli Institute for Astronomy and Astrophysics, Peking University, Yi He Yuan Lu 5, Haidian Qu, Beijing 100871, China}

\author[0000-0002-1838-4757]{Aniket Sanghi}
\affiliation{Department of Astronomy, The University of Texas at Austin,  2515 Speedway, Stop C1400 Austin, TX 78712, USA}

\author[0000-0002-7154-6065]{Gregory J. Herczeg}
\affiliation{Kavli Institute for Astronomy and Astrophysics, Peking University, Yi He Yuan Lu 5, Haidian Qu, Beijing 100871, China}

\author[0000-0001-9811-568X]{Adam L. Kraus}
\affiliation{Department of Astronomy, The University of Texas at Austin,  2515 Speedway, Stop C1400 Austin, TX 78712, USA}

\author[0000-0001-7258-770X]{Jaehan Bae}
\affiliation{Department of Astronomy, University of Florida, Gainesville, FL 32611, USA}

\author[0000-0002-7607-719X]{Feng Long}
\altaffiliation{NASA Hubble Fellowship Program Sagan Fellow}
\affiliation{Department of Planetary Science/Lunar and Planetary Laboratory, The University of Arizona, Tucson, AZ 85721, USA}

\author[0000-0002-7821-0695]{Katherine B. Follette}
\affiliation{Department of Physics and Astronomy, Amherst College, Amherst, MA 01003, USA}

\author[0000-0002-4479-8291]{Kimberly Ward-Duong}
\affiliation{Department of Astronomy, Smith College, Northampton, MA 01063, USA}

\author[0000-0003-3616-6822]{Zhaohuan Zhu}
\affiliation{Department of Physics and Astronomy, University of Neveda, Las Vegas,  NV 89154, USA}

\author[0000-0003-2646-3727]{Lauren I. Biddle}
\affiliation{Department of Astronomy, The University of Texas at Austin,  2515 Speedway, Stop C1400 Austin, TX 78712, USA}

\author[0000-0002-2167-8246]{Laird M. Close}
\affiliation{Department of Astronomy/Steward Observatory, University of Arizona, Tucson, AZ 85719, USA}

\author[0000-0003-4006-102X]{Lillian Yushu Jiang}
\affiliation{Department of Astronomy, The University of Texas at Austin,  2515 Speedway, Stop C1400 Austin, TX 78712, USA}

\author[0000-0002-4392-1446]{Ya-Lin Wu}
\affiliation{Department of Physics, National Taiwan Normal University, Taipei 116, Taiwan}
\affiliation{Center of Astronomy and Gravitation, National Taiwan Normal University, Taipei 116, Taiwan}

\begin{abstract}
The proposed protoplanet AB Aur b is a spatially concentrated emission source imaged in the mm-wavelength disk gap of the Herbig Ae/Be star AB Aur. Its near-infrared spectrum and absence of strong polarized light have been interpreted as evidence supporting the protoplanet interpretation. However, the complex scattered light structures in the AB Aur disk pose challenges in resolving the emission source and interpreting the true nature of AB Aur b. We present new images of the AB Aur system obtained using the Hubble Space Telescope Wide Field Camera~3 in the ultraviolet (UV) and optical bands. AB Aur b and the known disk spirals are recovered in the F336W, F410M, and F645N bands. The spectral energy distribution of AB Aur b shows absorption in the Balmer jump, mimicking those of early-type stars. By comparing the colors of AB Aur b to those of the host star, the disk spirals, and predictions from scattered light and self-luminous models, we find that the emission from AB Aur b is inconsistent with planetary photospheric or accretion shock models. Instead, it is consistent with those measured in the circumstellar disks that trace scattered light. We conclude that the UV and visible emission from AB Aur b does not necessitate the presence of a protoplanet. We synthesize observational constraints on AB Aur b and discuss inconsistent interpretations of AB Aur b among different datasets. Considering the significance of the AB Aur b discovery, we advocate for further observational evidence to verify its planetary nature.
\end{abstract}

\section{Introduction}

Directly imaged planets within protoplanetary disks hold valuable clues to understanding planet formation. They are hypothesized to be responsible for observed morphological and kinematic substructures in protoplanetary disks \citep[e.g.,][]{Zhang2018, Teague2018}. By imaging these planets, we can directly test theories of planet-disk interaction \citep[e.g.,][]{Keppler2018, Bae2019}, uncover mass-accretion processes \citep[e.g.,][]{Wagner2018,Marleau2022}, and constrain the timescales of giant planet formation \citep[e.g.,][]{Haffert2019,Choksi2023}. Additionally, their atmospheric composition and luminosity provide insights into initial states of planetary atmospheric and thermal evolution \citep[e.g.,][]{Stolker2020}. However, imaging protoplanets is challenging and prone to false detections. False-positive signals from image processing procedures \citep[e.g.,][]{Follette2017, Rameau2017} or stellar light scattered by the circumstellar disk \citep[e.g.,][]{Blakely2022} can mimic a pointlike source and confuse the identification of a planet. The discussion over the nature of AB Aurigae b (AB Aur b, reported by \citealt{Currie2022} as a protoplanet discovery) exemplifies both the excitement and the significant challenges associated with detecting protoplanets in high-contrast imaging data.

AB Aur is a remarkable system. The $\la5$\,Myr old, $2.4\pm0.2\,M_\odot$ \citep{Dewarf2003} Herbig Ae/Be star hosts a complex protoplanetary disk. Its ALMA 1.3~mm dust continuum image revealed a giant ($R=120$~au) inner cavity \citep{Tang2012,Francis2020}. Within the mm-dust cavity, multiple spiral arms have been identified by both infrared (IR) scattered light \citep[e.g.,][]{Hashimoto2011,Boccaletti2020,Jorquera2022,Betti2022} and $^{12}$CO gas emission \citep{Tang2017}, which point to disturbances from giant planets as a possible origin \citep[e.g.,][]{Oppenheimer2008,Tang2017,Boccaletti2020}. Using Subaru/SCExAO and Hubble Space Telescope (HST) high-contrast imaging observations, \citet{Currie2022} identified a spatially resolved clump that appears to be co-moving with the host star over a nearly fifteen-year baseline. These authors interpreted the clump as an embedded protoplanet and named it ``AB Aur b.'' If confirmed, AB Aur b offers a unique probe to a planet formation stage where the forming planet and its natal proplanetary disk interact.

Because AB Aur b is embedded in a complex disk, contamination due to scattered stellar light needs to be carefully removed to isolate and verify the planetary signal.
\citet{Currie2022} ruled out image post-processing artifacts as the origin of AB Aur b. They forward modeled the disk image and demonstrated that the point spread function (PSF) subtraction procedure did not produce an emission peak. \citet{Currie2022} also argued that the observed infrared (IR) emission from AB Aur b was not purely scattered stellar light due to the lack of spatially concentrated emission in the polarized light images. They reproduced the observed images with an embedded-protoplanet model in which the emission from the protoplanet is scattered off by marginally optically thin dust. Furthermore, \citet{Currie2022} showed that the near-IR spectral energy distribution (SED) of AB Aur b was bluer than those measured from several circumstellar disk locations. The spectral difference supported a self-luminous source at the AB Aur b location. 

Nevertheless, the embedded planet model cannot fit all observations of AB Aur b. Molecular absorption bands, which are hallmarks of planetary photospheric emission, have not been clearly detected in AB Aur b's IR spectrum \citep{Currie2022}. The optical continuum photometry measured by the HST Space Telescope Imaging Spectrograph (STIS) coronagraphic instrument (one single point covering 0.2 to 1.2 \micron{}) is overluminous compared to a planetary photospheric model \citep{Currie2022}. Accretion that produces hydrogen continuum emission is cited as a solution for elevating the optical continuum \citep{Currie2022}. However, a strong H$\alpha$ line that is often observed in accreting planets \citep[e.g.,][]{Wagner2018,Haffert2019,Hashimoto2020} and predicted by planetary accretion shock models \citep{Aoyama2018,Thanathibodee2019} is not detected in narrowband H$\alpha$ images \citep{Currie2022,Zhou2022}. Instead, the observed H$\alpha$ excess is consistent with the H$\alpha$ emission from accretion onto the host star and a scenario where the light source is dust scattering \citep{Zhou2022}.

The UV and optical colors of AB Aur b can provide another perspective to clarify the nature of this puzzling source. Possible models of AB Aur b predict vastly different SEDs at short wavelengths. In the scattered light scenario, the SED should resemble that of the host star with a peak at 0.4\,\micron{} and a dramatic drop in flux below the Balmer break at 0.36\,\micron{}. A self-luminous planetary photosphere should be bright in the IR and have red optical and UV colors. In contrast, an accreting planet should exhibit optical and UV excess flux and blue colors \citep[e.g.,][]{Zhu2015}. 

In this paper, we present new HST Wide Field Camera~3 (WFC3) images of AB Aur b taken in three bands covering 0.3 to 0.65\,\micron{} and identify the emission source of AB Aur b based on its observed colors. We describe the observations and data reduction procedures in Section~\ref{sec:2}, detail our immediate observational findings in Section~\ref{sec:3}, and synthesize all available constraints on AB Aur b in Section~\ref{sec:4}. We conclude with a summary in Section~\ref{sec:5}.

\section{Observations and data reduction}
\label{sec:2}

\begin{figure*}[!th]
    \centering
    \includegraphics[width=\textwidth]{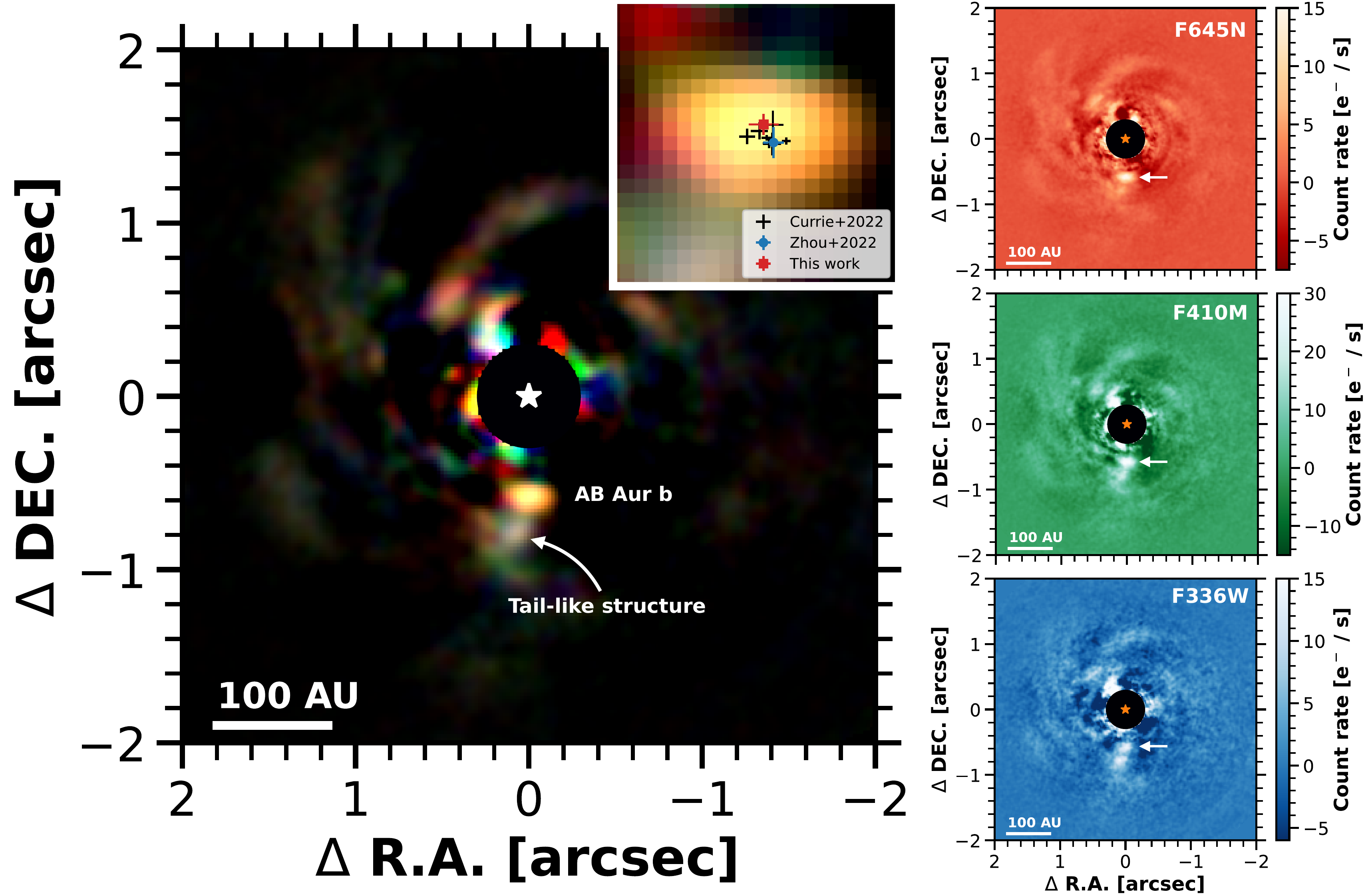}
    \caption{Primary subtracted images in RGB color (left), as well as in the F645N (R), F410M (G), and the F336W (B) bands (right). These images show a $2''{\times}2''$ region centered at AB Aur with north up and east pointing to the left. The colorbars are linearly stretched. In the right panels, arrows point to the AB Aur b feature. Excess flux at the position of AB Aur b is detected in all bands. Other prominent sources include several previously known spiral arms and a tail-like structure to the southeast of AB Aur b. The cut-out at the upper-right corner of the RGB image zooms into a $0''.4{\times}0''.4$ region centered at AB Aur b. The position of AB Aur b measured in this work (red) and reported in previous studies (\citealt{Currie2022}, black; \citealt{Zhou2022}, blue)  are over-plotted for comparison. Astrometry from this work does not follow the increasing PA trend predicted by the best-fitting Keplerian orbit in \citet{Currie2022}.}
    \label{fig:images}
\end{figure*}

AB Aur was imaged by HST/WFC3/UVIS2 from 2023-02-14 to 2023-02-16 UT in the F336W, F410M, and F645N bands for a total of nine HST orbits. These three filters cover the blue side of the Balmer jump, the blue optical continuum where an A0V SED peaks, and the red optical continuum near the H$\alpha$ line. We note that the F645N filter ($\lambda_\mathrm{cen}=6454\,$\AA, $\mathrm{FWHM}=84\,$\AA) is narrow enough to completely exclude the H$\alpha$ line. The images were taken with the \texttt{C512C} subarray ($512\times 512$ pixels, pixel scale=40~mas) that has a field of view of $20.5\times20.5$\,arcsec$^2$. To facilitate angular differential imaging (ADI, \citealt{Marois2006}), we positioned HST at five distinct roll angles, achieving a maximum field rotation of 37.3 degrees. In each orbit, the telescope cycled through four dithering positions that differed by 0.5 pixels and formed a box pattern.  At each dithering position, four 0.48\,s F336W, four 0.48\,s F410M, and three 2.0\,s F645N exposures were taken. In total, we obtained 144 F336W, 144 F410M, and 108 F645N frames.

Data reduction procedures are identical to those detailed in \citet{Zhou2021,Zhou2022}. We start by examining \texttt{flc}\footnote{The \texttt{flc} files are WFC3/UVIS calibrated exposures including CTE correction, see \href{https://hst-docs.stsci.edu/wfc3dhb/chapter-2-wfc3-data-structure/2-1-types-of-wfc3-files}{WFC3 Data Handbook}.} images by eye and identify spurious pixels and cosmic rays. These pixels are replaced by bilinear interpolations of their neighbors. We then combine each set of four dithered images into one Nyquist sampled ($\mbox{pixel scale} =20\,\mathrm{mas}$) image using the Fourier interlacing algorithm \citep{Lauer1999}. We use the \texttt{aperture\_photometry} function in the \texttt{photutils} package \citep{larry_bradley_2022_6825092} to perform aperture photometry for the host star. The aperture radius is 25 pixels (500\,mas) and the result is aperture corrected based on the UVIS2 encircled energy table \footnote{The UVIS2 encircled energy table is available at \url{https://www.stsci.edu/hst/instrumentation/wfc3/data-analysis/photometric-calibration/uvis-encircled-energy}}.

We then model and subtract the primary star PSF within a $r=2.5''$ circle of the star. The optimal PSF model is constructed by linearly combining the ADI images using the Karhunen-Lo\'eve Imaging Processing (KLIP) method implemented in the \texttt{pyKLIP} package \citep{Soummer2012,Wang2015}. We set the minimum field rotation difference between the target and PSF images to 25 degrees to minimize the impact of self-subtraction. At the separation of AB Aur b ($0''.6$, 30\,pixels), this angular difference translates to 13 pixels or $3.7{\times}\mathrm{FHWM}$ of the WFC3/UVIS2 PSF. Between $r_\mathrm{in}=0.1''$ and $r_\mathrm{out}=2.5''$, the images are split into seven concentric annuli, and the KLIP reduction is applied to each annulus independently. The inner three annuli has a narrower width ($\Delta r=0.2''$) than the outer four. The adopted number of KL modes is 30. Figure~\ref{fig:images} shows the primary-subtracted images in all three bands, as well as the combined three-color image (red: F645N, green: F410M, and blue: F336W). 

\begin{figure*}[!t]
    \centering
    \includegraphics[width=\textwidth]{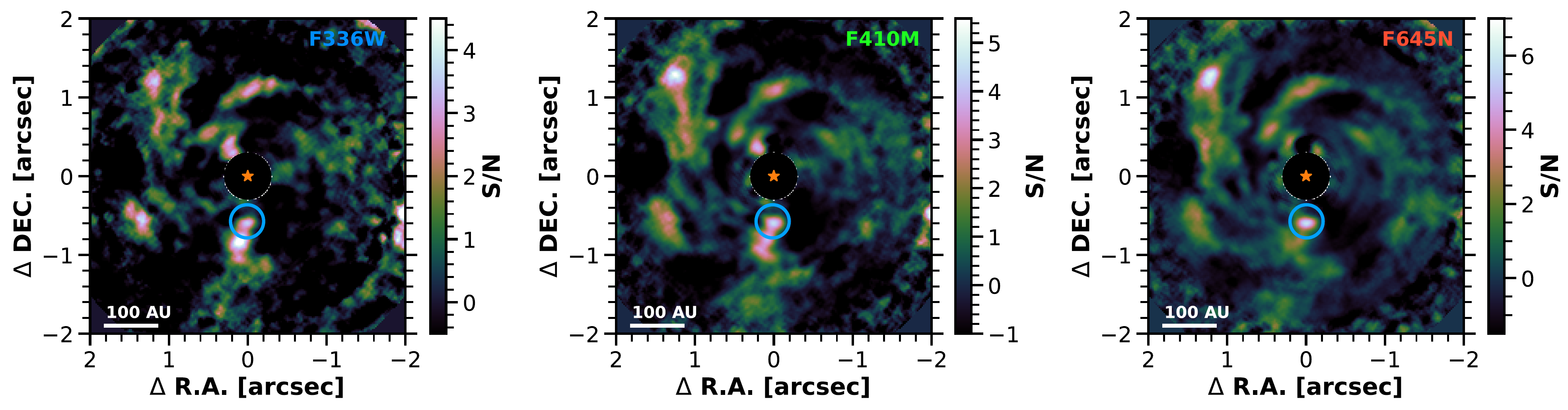}
    \caption{The S/N maps in the F336W, F410M, and F645N bands. Excess flux at the expected position of AB Aur b is consistently detected in all three bands. Circles mark the AB Aur b detections. These maps also reveal several other $\mathrm{S/N} > 3$ sources that align with the spiral sturctures observed in scattered light images \citep[e.g.,][]{Fukagawa2004,Boccaletti2020}. An extended source to the southeast of AB Aur b is detected at high significance, especially in the F336W and F410M bands.}\label{fig:snr}
\end{figure*}

We derive signal-to-noise (S/N) maps by dividing the primary-subtracted images over the noise maps. The noise maps are derived from the primary-subtracted image and dominated by correlated noise caused by residual speckles and flux from the disk. We assume the noise map to be azimuthally symmetric and use the standard deviation of the aperture-integrated flux of all non-overlapping apertures ($r=70\,\mathrm{mas}$, $1{\times}\mathrm{FWHM}$ at 0.4\,\micron{}\footnote{We note that the FWHM of the WFC3/UVIS PSF stays approximately constant at wavelengths between 0.4 and 0.7~\micron{}, see \href{https://hst-docs.stsci.edu/wfc3ihb/chapter-6-uvis-imaging-with-wfc3/6-6-uvis-optical-performance}{the WFC3 Instrument Handbook}.}) to estimate the correlated noise. Figure~\ref{fig:snr} presents the S/N maps in all three bands. Excess flux at the approximate position of AB Aur b as well as the locations of several previously identified disk spirals is detected in all three bands (Figures~\ref{fig:images} and \ref{fig:snr}). Notably, a  tail-like feature is detected near AB Aur b to its southeast. This feature is particularly prominent in the F336W and F410M bands.

\begin{figure}[!th]
    \centering
    \includegraphics[width=\columnwidth]{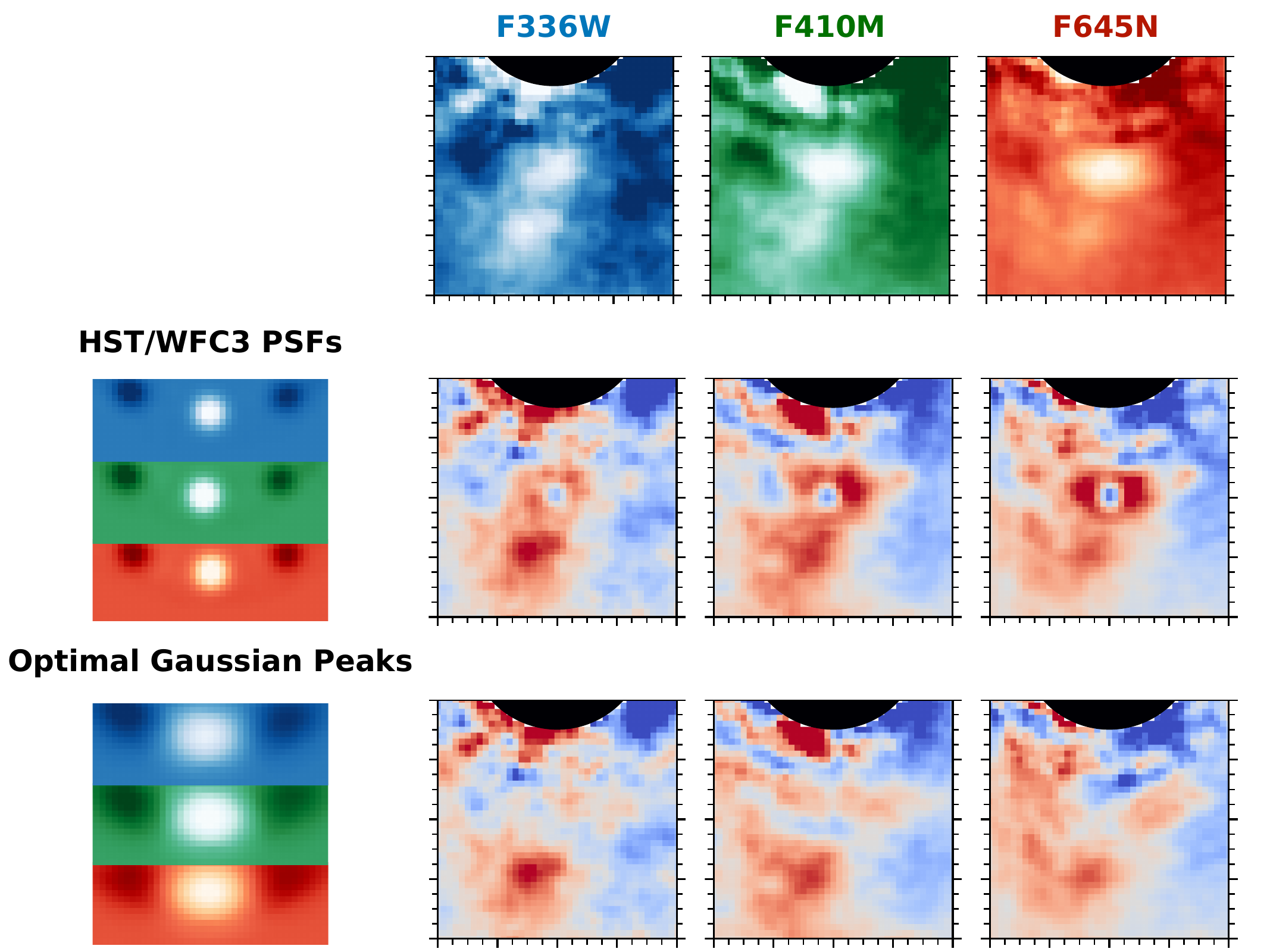}
    \caption{Constraining photometry, astrometry, and source morphology with KLIP-FM. The forward-modeled ADI images should contain no flux from the companion source when the model fits. For images shown in the middle row, HST PSFs are injected, resulting in poor fits and significant residual flux near AB Aur b. For images shown in the bottom panel, optimal 2D Gaussian peaks are injected, eliminating excess flux. The source to the southeast of AB Aur b is not considered in the forward-modeling process, and it remains visible in all images.}
    \label{fig:fm}
\end{figure}

\newcommand{\fwhm}{\ensuremath{\mathrm{FHWM}}\xspace}
The AB Aur b feature is spatially extended and broader than the WFC3 PSF (Figure~\ref{fig:fm}), particularly in the azimuthal direction. We characterize the source morphology by fitting a two-dimensional Gaussian peak to the primary-subtracted images. Because the companion PA is consistent with $180^\circ$, the $x$ and $y$ axes align with azimuthal and radial directions, respectively. The best-fitting sizes are ($\sigma_x, \sigma_y$) = (74, 68), (86, 70), and (90, 54)~mas in the F336W, F410M, and F645N images, respectively. The average sizes are $\sigma_x=84$\,mas and $\sigma_y=64$\,mas. After deconvolving with the WFC3 PSF ($\sigma=30$~mas), we find ($\sigma_{x0}, \sigma_{y0}$) = (68, 61), (81, 63) and (85, 45)\,mas in the three bands and average sizes of $\sigma_{x0}=78$\,mas and $\sigma_{y0}=56$\,mas. The average projected physical sizes are 12\,au and 8.7\,au in the azimuthal and radial directions, respectively.  These sizes are slightly larger than those found in the Subaru/CHARIS data and consistent with those found in the HST/STIS images \citep{Currie2022}.

We measure the photometry and astrometry of AB Aur b using the KLIP forward modeling method \citep{Pueyo2016}. We assume the best-fitting Gaussian peak as the source model for each band and use the \texttt{fm} module of the \texttt{pyKLIP} package to apply ADI distortion to the model. The position and amplitude of the peaks are optimized using Markov Chain Monte Carlos posterior sampling. The medians of the posterior distributions are adopted as the source's flux and position, and the standard deviations are adopted as uncertainties. As shown in Figure~\ref{fig:fm}, the optimized source models fit the AB Aur b observations well, allowing accurate photometric and astrometric measurements.

We also performed aperture photometry for AB Aur b and the spiral structures at locations where S/N is reasonably high. To locate the aperture centers of the spiral arm sources, we first identify patches where the $\mathrm{S/N}=2$ to 4 in the three bands. These patches sample the distributions of surface density and scattering phase functions within the AB Aur disk. We then use the \texttt{centroid\_com} function in the \texttt{photutils} package to find the centroid for all patches. We use $r=70$\,mas ($1{\times}$FWHM) apertures. The aperture-integrated flux is corrected for KLIP throughput (estimated by injecting and recovering point sources) and the finite aperture size (using the WFC3/UVIS2 encircled energy curves). %We note that aperture photometry implicitly assumes point sources that do not describe AB Aur b or spiral arms. Nevertheless, the results provide informative estimates of the colors of these sources.

\section{Results}
\label{sec:3}

We detect excess flux at the expected position of AB Aur b in all three bands. The detection S/N levels are \SNRUV, \SNRG, and \SNRR in the F336W, F410M, and F645N bands, respectively (Figures~\ref{fig:images}, \ref{fig:snr}). The position angle and separation of AB Aur b are $181^\circ.6\pm4^\circ.8$ and $574\pm35$\,mas in F336W, $182^\circ.6\pm3^\circ.9$ and $573\pm13$\,mas in F410M, and $180^\circ.2\pm3^\circ.0$ and $577\pm22$\,mas in F645N relative to the host. Taking the average of three bands using inverse-variance weighting yield $\mathrm{PA}=181^\circ.2\pm2^\circ.1$ and $\mathrm{separation}=574\pm15$\,mas. The astrometry is consistent with the latest published results within $1\sigma$ (epoch 2020-2021: \citealt{Currie2022}, epoch 2022: \citealt{Zhou2022}). However, the PA does not follow the increasing trend predicted by the best-fitting Keplerian orbit in \citet{Currie2022} (which is notably eccentric with $e\sim 0.2$ to 0.6) and deviates from the predicted PA ($183^\circ.9$)  by $1.3\sigma$.

\begin{figure*}[!th]
\includegraphics[width=\textwidth]{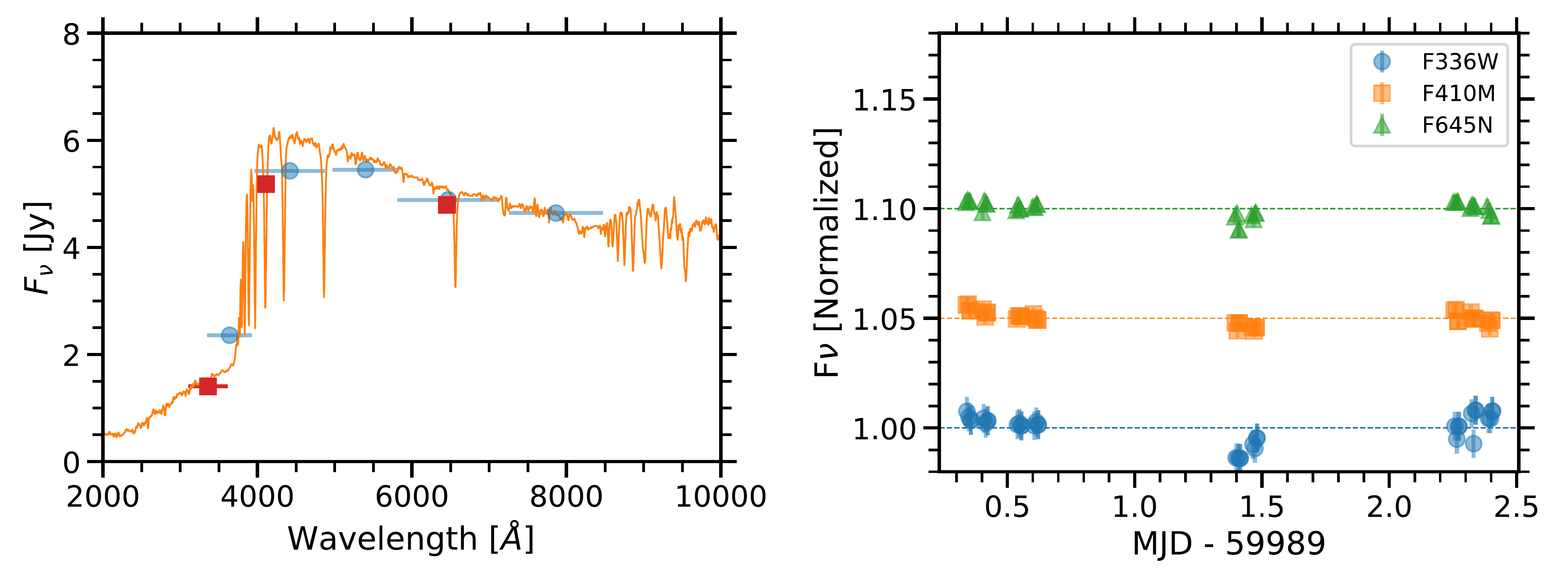}
\caption{WFC3 photometry of the host star, AB Aur. \textit{Left:} the F336W, F410M, and F645N flux densities of AB Aur A (red squares) are compared with archival photometry (blue circles, U/B/V/R/I bands, \citealt{Lazareff2017}) as well as a slightly reddened A0V spectrum (shown in orange, $A_V=0.4\,$mag). The consistency verifies the absolute photometric calibration of our observations. \emph{Right:} the time-resolved photometry of AB Aur A in the three observed bands. Offsets are applied to the F410M and F645N data for clarity. No significant variability is detected during our observations.}\label{fig:hostflux}
\end{figure*}

Table~\ref{tab:phot} lists photometry for AB Aur A and b, as well as their $\mathrm{F336W}-\mathrm{F410M}$ and $\mathrm{F410M}-\mathrm{F645N}$ colors. The WFC3 photometry together with optical photometry from the literature of AB Aur A \citep{Lazareff2017} is consistent with a slightly reddened A0V ($A_V=0.4$\,mag) spectrum (Figure~\ref{fig:hostflux}). This independently validates the absolute photometric calibration of our observations. No significant variability in the host star is detected within the 2.5\,days over which the HST observations were conducted (Figure~\ref{fig:hostflux}). The standard deviations of the time-resolved photometry are 0.65\%, 0.30\%, and 0.31\% in the F336W, F410M, and F645N bands, respectively.
\begin{deluxetable}{lcc}
\tablecaption{Photometry of AB Aur A\&b} \label{tab:phot}
\tablewidth{0pt}
\tablehead{
\colhead{Band} & \colhead{$f_\nu$ (mJy)} &\colhead{magnitude}
}
\startdata
\multicolumn{3}{c}{AB Aur A} \\
\hline
F336W & $1400\pm9$ & $7.370\pm0.007$\\ 
F410M & $5180\pm16$ & $7.288\pm0.003$\\
F645N & $4790\pm15$ & $7.004 \pm0.003$\\
F336W$-$F410M & $\cdots$ &$0.082\pm0.007$\\
F410M$-$F645N & $\cdots$ &$0.285\pm0.004$\\
\hline
\multicolumn{3}{c}{AB Aur b (aperture photometry)} \\
\hline
F336W & $0.121\pm0.031$ & $17.53\pm0.28$\\ 
F410M & $1.08\pm0.24$ & $16.49\pm0.24$\\
F645N & $1.36\pm0.25$ & $15.87 \pm0.20$\\
F336W$-$F410M & $\cdots$ &$1.04\pm0.37$\\
F410M$-$F645N & $\cdots$ &$0.62\pm0.31$\\
\hline
\multicolumn{3}{c}{AB Aur b (forward modeling)} \\
\hline
F336W & $0.141\pm0.030$ & $17.36\pm0.24$\\ 
F410M & $1.13\pm0.18$ & $16.44\pm0.18$\\
F645N & $1.54\pm0.19$ & $15.74 \pm0.14$\\
F336W$-$F410M & $\cdots$ &$0.92\pm0.29$\\
F410M$-$F645N & $\cdots$ &$0.70\pm0.22$\\
\enddata
\end{deluxetable}

\begin{figure*}[!th]
    \centering
    \includegraphics[width=\textwidth]{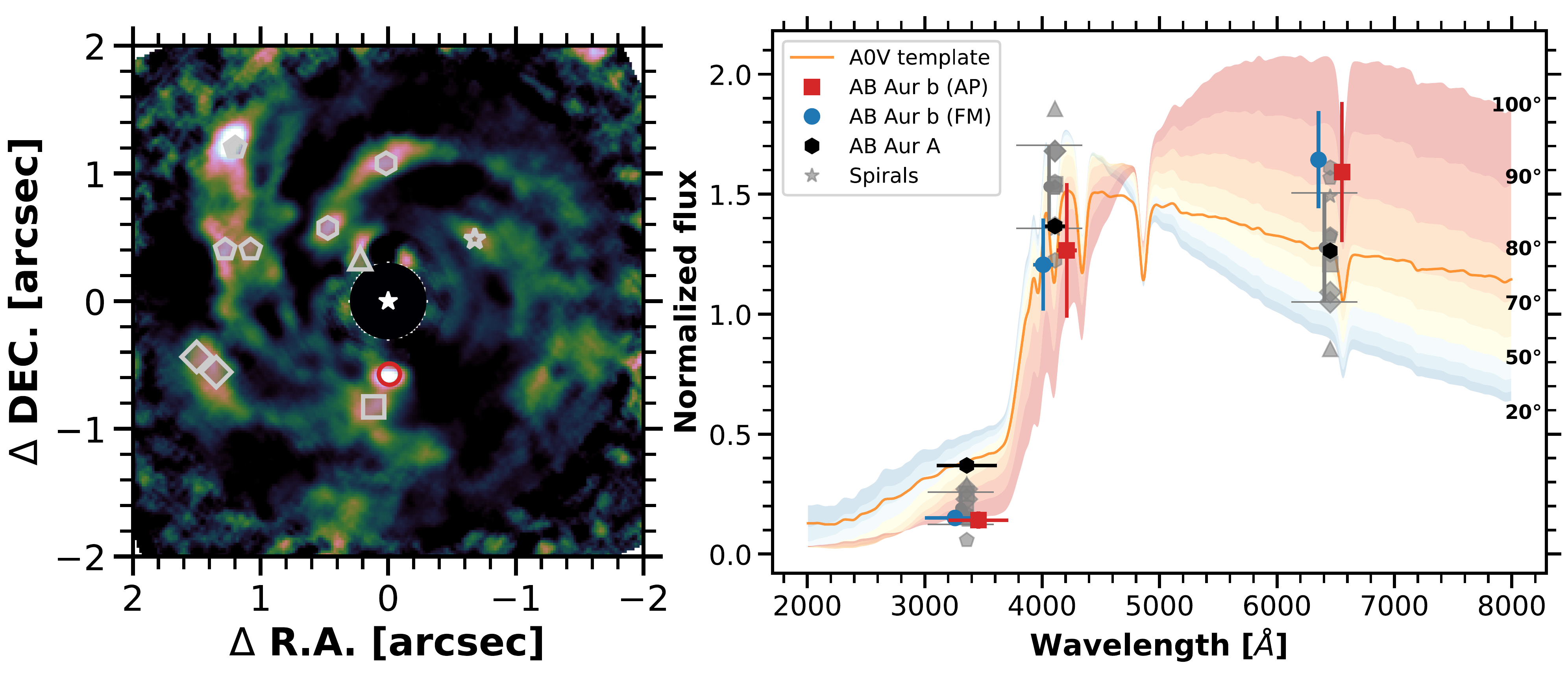}
    \caption{Measuring the colors of scattered light disk features. \textit{Left:} the markers show the locations where apertures are placed to measure photometry of disk spirals. Apertures cover the same spiral share marker symbols. \textit{Right:} the comparison of SEDs of AB Aur (black hexagon), AB Aur b (blue circle, forward modeling; red square: aperture photometry), and the spiral disk (various gray markers). Each SED is normalized by dividing its averaged flux among the three band. The gray vertical bars show the $1\sigma$ standard deviation of the normalized flux measured in the spirals. A scaled A0V spectral template and the scattered light spectra (0.1\,\micron{} spherical particle, Mie-scattering) are over-plotted for comparison. The colors (blue to red) of the lines represent scattering angles ($20^\circ$, forward scattering to $100^\circ$, backward scattering).}
    \label{fig:color-image}
\end{figure*}

Despite AB Aur b's deviation from a point source (Figure~\ref{fig:fm}), forward modeling and aperture photometry yield consistent flux density measurements. Using forward modeling, we find $f_\nu$ of $0.141\pm0.030$\,mJy in F336W, $1.13\pm0.18$\,mJy in F410M, and $1.54\pm0.19$\,mJy in F645N; using aperture photometry, we find the respective flux densities are $f_\nu=0.121\pm0.031$, $1.08\pm0.24$, and $1.36\pm0.25$\,mJy. AB Aur b's SED is more similar to that of an early-type star than a substellar object (Figure~\ref{fig:color-image}). UV excess flux at the Balmer jump ($\lambda=3645\,$\AA) that traces accretion onto low-mass stars and substellar objects is not detected in AB Aur b.

The new HST images also reveal several spiral structures that have been previously identified in scattered light images of the AB Aur disk (marked by different symbols in Figure~\ref{fig:color-image}, \citealt{Fukagawa2004,Boccaletti2020}). Their brightness and colors are empirical references for scattered light disk features. The detection S/Ns are low ($\mathrm{S/N}\sim2$ to 4), and the spiral shapes can be distorted by ADI \citep{Boccaletti2020}. We present their SEDs and colors that are determined by aperture photometry in Figures~\ref{fig:color-image} and \ref{fig:color-model}. The color spread is mostly measurement noise, although differences in scattering phase function can also contribute. 
Overall, the spirals share a consistent SED shape that peaks at around 0.4\,\micron{} and drop significantly in the UV. On average, these structures have a F336W${-}$F410M color of $1.0\pm0.4$~mag and a F410M${-}$F645N color of $0.16\pm0.32$~mag. Their colors do not deviate from that of AB Aur b beyond their mutual uncertainties.

One of the most interesting features newly identified in these observations is a tail-like structure adjacent to AB Aur b, located at a separation of $0.84''$ and a PA of $172.1^\circ$ relative to the host star and positioned at a separation of $0.28''$ from AB Aur b. This structure lies outside the aperture or the forward-modeling optimization region, and consequently has no impact on the photometry of AB Aur b. This feature has not been identified in previous observations but it is detected with $\mathrm{S/N} > 3$  in the F336W and F410M bands. Its flux densities are $0.086\pm0.016$, $0.53\pm0.14$, and $0.42\pm0.15$\,mJy in the F336W, F410M, and F645N bands, which translate to a F336W${-}$F410M color of $0.63\pm0.35$ mag and a F410M${-}$F645N color of $0.11\pm0.48$\,mag. The tail is bluer than AB Aur b by 0.41 and 0.51 mag in F336W${-}$F410M and F410M${-}$F645N, but these differences do not exceed their mutual uncertainties. The colors of this tail-like structure are also within $1\sigma$ spread of the colors of the spirals, so it is most likely caused by scattered light. 

\section{Discussion}
\label{sec:4}

\subsection{The optical continuum of AB Aur b is brighter in HST/WFC3 than HST/STIS.}

We find brighter optical continuum flux from AB Aur b than that measured by HST/STIS \citep{Currie2022}. Considering the throughput difference between the WFC3 and STIS filters, we translate the WFC3 photometry to a $f_\nu=0.88$~mJy flux in the HST/STIS bandpass by assuming the true SED has a A0 shape. This flux density is significantly higher than that reported by \citet[0.48\,mJy]{Currie2022}. The flux difference may indicate variability, which could be caused by variations in the incident stellar flux, the size and shape of a dust clump (assuming the optical band flux is scattered light), or luminosity of the companion (assuming it is self-luminous). Variability has been observed in both accreting planetary mass objects \citep[e.g.,][]{Demars2023} and scattered light images of transition disks \citep[e.g.,][]{Pinilla2015}.

This flux difference can also arise from systematic errors in flux calibration. The photometric methods are slightly different between \citet{Currie2022} for the STIS data and this work for the WFC3 data. The former fitted a Gaussian peak model and we performed forward modeling. We have verified the absolute flux calibration of the WFC3/UVIS instrument by demonstrating that the observed host stellar flux is consistent with the literature photometry (Figure~\ref{fig:hostflux}). We also obtain consistent photometry of AB Aur b using two distinct methods. There is a lack of cross-calibration of high-contrast photometry between the WFC3/UVIS and STIS instruments. For this discussion, we focus on using the WFC3/UVIS photometry to constrain the nature and properties of AB Aur b.

\subsection{The UV and optical flux from AB Aur b is most likely scattered light.}

% \begin{figure*}[t]
%     \centering
%     \includegraphics[width=1.0\textwidth]{SED.pdf}
%     \caption{The UV-visible-IR SED of AB Aur b. New data presented in this paper are shown in colored squares, and published data from \citep{Currie2022,Zhou2022} are shown in the black line and squares. For the new photometry, measurements by three methods are all included and plotted in different colors (blue: forward modeling, yellow: peak fitting), red: aperture photometry) to reflect the systematic uncertainties in absolute flux calibration between photometric methods and instruments. The peak-fitting results are plotted in a more prominent style because it is consistent with the one used to determine the visible band photometry from HST/STIS \citep{Currie2022}. Three representative models are shown in dashed lines: pink: accretion shock modeled by a hydrogen slab, red: scattered stellar light, green: self-luminous black body. Synthetic photometry in the F336W, F410M, and F645N bands are shown in circles with respective colors. The observations are most consistent with the scattered light model.}
%     \label{fig:SED}
% \end{figure*}

The UV-optical continuum flux observed at the AB Aur b may come from two sources: 1. a self-luminous and embedded planet; or 2. scattered light by substructures in the circumstellar disk. In the self-luminous scenario, the emission may originate from a photosphere, an accretion shock, or a combination of the two and then gets reprocessed by a dusty envelope. Here we derive the synthetic colors of these models and compare them with the observations.

\newcommand{\teff}{\ensuremath{T_\mathrm{eff}}\xspace}
\newcommand{\logg}{\ensuremath{\log g}\xspace}

\begin{figure*}[!t]
    \centering
    \includegraphics[width=\textwidth]{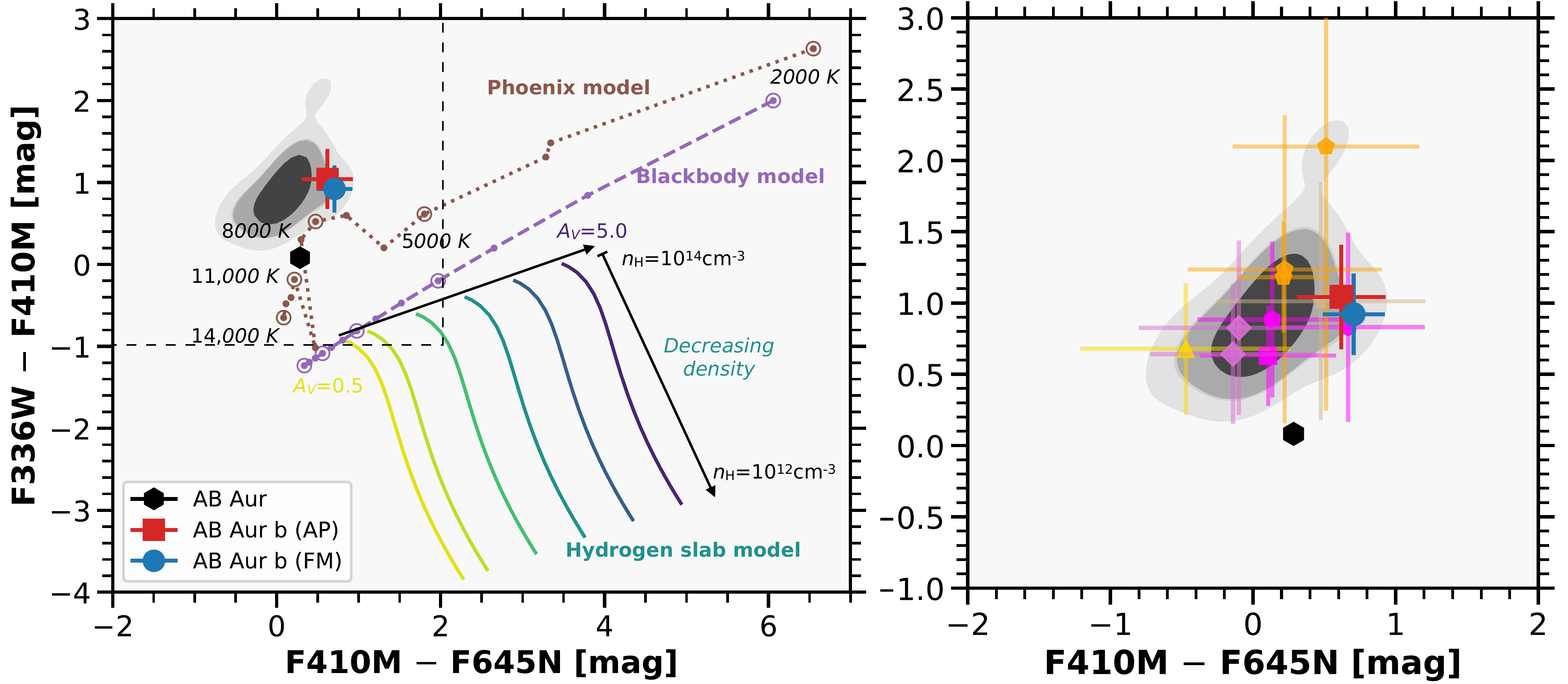}
    \caption{Color-color diagrams containing AB Aur, AB Aur b, and the circumstellar disk sources measured in the various apertures along spiral arms. The symbols are identical to those used in Figure~\ref{fig:color-image}. The gray contours show the 1, 2, and 3$\sigma$ kernel density estimates of the spiral sources. \textit{Left:} an extended color-color diagram to enclose possible self-luminous models, including photosphere (PHOENIX), blackbody, and accretion shock models. None of these models reproduces the observed color of AB Aur b. \textit{Right:} the color-color diagram zoomed into the vicinity of AB Aur b. Dots and errorbars represent the spiral sources.  The color of AB Aur b is consistent with disk sources that trace scattered stellar emission. The empirically estimated scattered light colors match the observed colors of AB Aur b much better than self-luminous models.}
    \label{fig:color-model}
\end{figure*}

We assume a blackbody for a simplified protoplanetary photospheric model. The model choice is justified by the lack of strong molecular features observed in PDS 70 b, c, and AB Aur b \citep[e.g.,][]{Stolker2020,Currie2022}. We consider blackbody temperatures ranging from 2000\,K to 14,000\,K, an extinction of $A_V=0.5$\,mag (the same as the host star extinction) and show the synthetic colors in Figure~\ref{fig:color-model}. The blackbody models fail to reproduce the observed colors. The cooler models ($\teff<5000$\,K) are redder in the F410M$-$F645N color, the hotter models ($\teff>5000$\,K) are bluer in the F336W$-$F410M color, and all blackbody colors differ from the observations of AB Aur b by more than two magnitudes. Adopting a greater extinction pushes the model further away from the observations in the color-color space. 

The more realistic PHOENIX model \citep{Allard2003} cannot reproduce the observed colors with temperatures consistent with a substellar photosphere, either. We experiment with models with  \logg of 4.5\,dex [cgs units], extinction of $A_V=0.5$\,mag, and  a range of  \teff from 2000\,K to 14,000\,K. Only when \teff is well within the stellar regime ($\teff\sim7000$\,K to 8000\,K) does the PHOENIX spectrum produces similar colors to those of AB Aur b. The lower \teff models have significantly redder F410M$-$F645N colors. We reject a planetary photosphere as the UV and optical emission source based on these comparisons.

A plane-parallel hydrogen slab model is adopted to estimate the colors of accretion shocks. These models reproduce the continuum emission observed in T Tauri stars, brown dwarfs, and planetary-mass objects \citep[e.g.,][]{Valenti1993,Herczeg2008,Zhou2014}. We adjust the hydrogen atom number density ($n_\mathrm{H}\in(10^{12}, 10^{14})\,\mathrm{cm^{-3}}$) to set the slab optical depth and size of the Balmer jump. At the low density end, the slab is optically thin and produce a pronouced UV flux jump; at the high density end, the slab is optically thick and the output SED regresses to a blackbody. We experiment with a wide range of extinction levels ($A_V=$\,0.5, 1, 2, 3, 4, and 5\,mag). The synthetic colors of these models are shown in Figure~\ref{fig:color-model}. Due to the presence of a Balmer jump, the low-extinction slab models are significantly bluer ($>2$\,mag) than AB Aur b in the $\mathrm{F336W}-\mathrm{F410M}$ color. The high-extinction models are significantly redder ($>2$~mag) in the $\mathrm{F410M}-\mathrm{F645N}$ color. The disagreement between the hydrogen slab models and the observations allows us to reject an accretion shock as the emission source. Linearly combining a photosphere and an accretion shock cannot reproduce the observed color either. In summary, the UV and optical emission observed in AB Aur b is not from a self-luminous source.

\begin{deluxetable*}{llc}
\tablewidth{\textwidth}
\tablecaption{Observations and interpretations of AB Aur b}\label{tab:obs}
\tablehead{
  \colhead{Observations} &
  \colhead{Interpretations} &
  \colhead{Refs}
}
\startdata
Non-detection in polarized light & AB Aur b cannot be a pure protoplanetary disk feature. & [1]\\
IR SED & AB Aur b is self-luminous. & [1]\\
HST/STIS photometry  & Observed flux may come from both the planetary photosphere and magnetospheric accretion. & [1]\\
H$\alpha$ excess & The H$\alpha$ detection can be attributed to scattered stellar light. & [1, 2]\\
UV and optical colors & The UV and optical flux is scattered stellar light. & [3]\\
&No self-luminous source is seen at short wavelengths. &
\enddata
\tablerefs{
[1] \citet{Currie2022};
[2] \citet{Zhou2022};
[3] This work.
}
\end{deluxetable*}

The consistent colors between AB Aur b and the disk spirals suggest that the UV and optical flux observed in AB Aur b is most likely scattered light. Several scattered light spectra derived using the Mie theory \citep{Bohren1983} are compared with the observations in Figure~\ref{fig:color-image}. In modeling the scattered light, we omit radiative transfer calculations and assume a single scattering scenario. The grains are spherical, effectively uniform in size, and have composition adopted to interpret disk images from the Disk Substructures at High Angular Resolution Project (DSHARP, \citealt{Andrews2018} and \citealt{Birnstiel2018}). The model with a $r=0.1$\,\micron{} grain size and a $70^\circ$ phase angle reproduces the SED of AB Aur b.  We have also experimented with other grain sizes and find that smaller grains (0.01\,\micron{}) make scattered light too blue, whereas larger grain (0.5\,\micron{} and 1\,\micron{}) produces scattered light that is too red. Notably, our model can be overly simplified as single scattering is insufficient in an optically thick disk where multiple scattering is likely occuring. Moreover, recent studies of scattered light images of debris disks have found that models based on Mie theory or distributions of hollow spheres \citep{Min2005} could not reproduce total intensity and polarized light observations consistently \citep{Ren2019,Arriaga2020}. Because of the modeling limitations, we emphasize that the empirical result --- color similarities between AB Aur b and the circumstellar disks --- is the most clear evidence for the scattered light origin of AB Aur b's UV and optical flux.

%Stellar light scattered by sub-\micron-sized grains shares the colors of AB Aur b. The consistent colors between AB Aur b and the disk spirals support this conclusion. We further examine this result by modeling the color of the stellar light using the Mie theory \citep{Bohren1983}. The grains are effectively uniform in size and have the DSHARP composition \citep{Birnstiel2018}. For simplicity, we omit radiative transfer calculations and assume the single scattering scenario. Under this assumption, scattering happens within the $\tau=1$ surface of the disk, and the scattered light intensity can be approximated as:
%\begin{equation}
%I_\lambda \propto I_{\lambda, 0} \eta_\lambda \Phi_\lambda(\theta),
%\end{equation}
%where $I_{\lambda, 0}$ is the incident intensity, $\eta_\lambda = \kappa_{\rm sca, \lambda}/\kappa_{\rm ext, \lambda}$ is the albedo, and $\Phi_\lambda(\theta)$ is the normalized scattering phase function, and  $\theta$ is the scattering angle. We calculate scattere light spectrum using four representative grain sizes (0.01\,\micron{}, 0.1\,\micron{}, 0.5\,\micron{}, and 1\,\micron{}) and find that the model with 0.1\,\micron{} dust and a $\theta\approx70^\circ$ scattering angle matches the observed colors of AB Aur b and the spirals the best (Figures~\ref{fig:color-image} and \ref{fig:color-model}). Smaller grains (0.01\,\micron{}) makes scattered light too blue, whereas larger grains (0.5\,\micron{} and 1\,\micron{}) produces scattered light that is too red.

To conclude, the UV and optical flux observed at the position of AB Aur b is likely scattered stellar emission. We rule out a bare planetary photosphere or an accretion shock at the planetary or disk surface as possible emission sources. A self-luminous planet may be deeply embedded in the circumstellar disk. In this scenario, its short-wavelength emission is significantly attenuated by its dust envelope and does not significantly contribute to the visible and UV wavelength flux (at least during this observation epoch).

\subsection{What is AB Aur b?}

Many observations have been collected for AB Aur b. We list the observational results presented in \citet{Currie2022}, \citet{Zhou2022}, and this work, as well as the most likely interpretations in Table~\ref{tab:obs}. Altogether these data do not point to a coherent conclusion as to whether AB Aur b is a self-luminous planet. We summarize the conflicting points below.

Based on the blue IR SED (relative to those measured at eight locations in the disk of AB Aur, see Figure S13 of \citealt{Currie2022}) and the lack of polarized light detection, \citet{Currie2022} concluded that AB Aur b is a self-luminous planet. This disagrees with what we find from the UV and optical data. The colors between self-luminous models and scattered light are more distinguishable in UV and optical bandpasses. In the IR, the scattered light source is the inner disk around the host star, which has similar colors as a cool planetary photosphere \citep{Furlan2006,Betti2022}. In the UV and optical bands, the scattered light source is dominated by the A-type stellar photosphere that differs dramatically from any plausible self-luminous planetary sources (Figure~\ref{fig:color-model}). Although we confidently conclude that the AB Aur b's UV and optical flux is scattered light, we emphasize that the scattered light model struggles to explain AB Aur b's low polarization ratios in the IR. Recent advances in dust modeling in circumstellar disks offer a possible pathway towards reproducing the total intensity and polarized light observations with scattered light models \citep{Tazaki2023}.

Based on the IR and visible SED, \citet{Currie2022} suggested a combination of a planetary photosphere and a magnetospheric accretion shock as a viable model. However, strong H$\alpha$ excess emission, which is observed in PDS 70 b\&c \citep[e.g.,][]{Haffert2019} and predicted by planetary accretion shock models \citep[e.g.,][]{Aoyama2018}, was not found in AB Aur b \citep{Currie2022,Zhou2022}. As we have shown in the previous subsection and Figure~\ref{fig:color-model}, the UV and optical colors of AB Aur b are inconsistent with accretion shock models. The overluminous HST/STIS photometry compared to the photospheric model best-fit to IR photometry may be attributed to the challenges in calibrating photometry for extended sources located within a complex disk, especially when combining measurements from multiple instruments.

\section{Summary}
\label{sec:5}

We present new HST high-contrast imaging observations of AB Aur, which hosts a complex transition disk and an embedded protoplanet candidate. Our findings are summarized as follows:

1. We observed AB Aur from 2023-02-14 to 2023-02-16 UT in the F336W, F410M, and F645N using HST/WFC3. After subtracting the primary star, we detect concentrated flux at the location of AB Aur b with S/N levels of \SNRUV, \SNRG, and \SNRR in the three bands. We have also detected previously known disk spiral structures at lower S/N. 

2. The AB Aur b detections are spatially extended, especially in the azimuthal direction. The source can be characterized by a two-dimensional Gaussian peak with $\sigma_x=78$\,mas and $\sigma_y=56$\,mas, after deconvolving the instrument PSF. The projected physical sizes are 12\,au and 8.7\,au in the azimuthal and radial directions. 

3. The astrometry and photometry of AB Aur b are measured using KLIP forward modeling.  The PA and separation of AB Aur b are $181^\circ.6\pm4^\circ.8$ and $574\pm35$ in F336W, $182^\circ.6\pm3^\circ.9$ and $573\pm13$, $180^\circ.9\pm1^\circ.6$ and $569\pm28$ in F410M, and $180^\circ.2\pm3^\circ.0$ and $577\pm22$\,mas in F645N relative to the host. The average values are $\mathrm{PA}=181^\circ.2\pm2^\circ.1$ and $\mathrm{separation}=574\pm15$\,mas. The flux densities are $0.137\pm0.030$, $1.08\pm0.24$, and $1.36\pm0.25$~mJy in the F336W, F410M, and F645N bands, respectively. Aperture photometry is also performed on AB Aur b and is consistent with these measurements (Table~\ref{tab:phot}).

4. Several previously known spiral structures in AB Aur's disk are detected. We identify regions in the spiral structures with elevated S/N and determine their brightness and colors using aperture photometry. Their ensembled colors, which serve as empirical references of scattered light colors, are consistent with the UV-optical colors of AB Aur b.

5. The observed UV-optical colors of AB Aur b are compared with a wide range of plausible models, including accretion shocks, planetary photospheres, and scattered light. We reject self-luminous models as possible sources of the UV-optical emission.
% Scattered light is the most likely origin of the observed UV and optical emission.

6. We do not detect any accretion-related UV-excess emission. This rules out accretion as an origin of AB Aur b's blue optical color.

There remain considerable uncertainties in the interpretation of AB Aur b. We advocate for future efforts to conclusively determine its nature. The photometry of AB Aur b measured by ground- and space-based instruments needs to be calibrated for constructing and analyzing the full UV, optical, and IR SED. Given the complexities in high-contrast imaging, this task is not trivial. A valid model of AB Aur b needs to reconcile the conflicts between the IR observations that have been interpreted as a self-luminous source and UV-optical observations that are dominated by scattered light.

\facilities{Hubble Space Telescope}

\software{astropy \citep{AstropyCollaboration2022}, numpy \citep{harris2020array}, scipy \citep{2020SciPy-NMeth}, matplotlib \citep{Hunter:2007}, seaborn \citep{Waskom2021}, photutils \citep{larry_bradley_2022_6825092}, pysynphot \citep{STScIDevelopmentTeam2013}, pyKLIP \citep{Wang2015}, emcee \citep{Foreman-Mackey2013}}

\vspace{2em}
We thank the anonymous referee for a careful and constructive report that improves this manuscript. We acknowledge excellent observing support from STScI staffs P. McCullough and T. Royle. We thank Dr. Andrea Isella for helpful discussion. Y.Z. acknowledges support from the Heising-Simons Foundation 51 Pegasi b Fellowship. B.P.B. acknowledges support from the National Science Foundation grant AST-1909209, NASA Exoplanet Research Program grant 20-XRP20\_2-0119, and the Alfred P. Sloan Foundation. G.J.H. is supported by general grant 12173003 from the National Natural Science Foundation of China. Support for F.L. was provided by NASA through the NASA Hubble Fellowship grant \#HST-HF2-51512.001-A awarded by the Space Telescope Science Institute, which is operated by the Association of Universities for Research in Astronomy, Incorporated, under NASA contract NAS5-26555. The observations and data analysis works were supported by program HST-GO-17280. Supports  for  Program  numbers  HST-GO-17280 were provided by NASA through a grantfrom the Space Telescope Science Institute, which is operated by the Association of Universities for Research in Astronomy, Incorporated, under NASA contract NAS5-26555.

\bibliographystyle{aasjournal}
\bibliography{library}

\end{document}